\documentclass[conference]{IEEEtran}
\pdfoutput=1

\usepackage[export]{adjust box}
\usepackage{pgf}
\usepackage{booktabs}
\usepackage{balance}
\usepackage{paralist}
\usepackage{amsmath}
\usepackage[normalem]{ulem}
\usepackage{color}

\usepackage{url}
\usepackage{graphicx}
\usepackage{float}
\usepackage{array}
\usepackage{multirow}
\usepackage{setspace}
\usepackage{textcomp}

\usepackage{float}
\usepackage[caption = false]{subfig}

\usepackage{dblfloatfix}

\usepackage{xcolor,colortbl}

\definecolor{Gray}{gray}{0.85}
\newcolumntype{a}{>{\columncolor{Gray}}c}

\newcolumntype{C}[1]{>{\centering\arraybackslash}m{#1}}
\usepackage[normalem]{ulem}
\useunder{\uline}{\ul}{}
\usepackage{booktabs} 
\usepackage{color}

\newcommand{\spara}[1]{\smallskip\noindent{\bf #1}}

\usepackage{fancyhdr} 
\usepackage{kantlipsum} 
\fancypagestyle{plain}{ 
\fancyhf{} 
\fancyhead[C]{Conference on \LaTeX} 
 
} 
\usepackage{eso-pic} 
\begin{document}

\AddToShipoutPictureBG*{ 
\AtPageUpperLeft{ 
\setlength\unitlength{1in} 
\hspace*{\dimexpr0.5\paperwidth\relax}
\makebox(0,-0.75)[c]{\textbf{2018 IEEE/WIC/ACM International Conference on Web Intelligence (WI'18)}}}} 
\title{Distributional Semantics Approach to Detect Intent in Twitter Conversations on Sexual Assaults}
\author{\IEEEauthorblockN{Rahul Pandey}
\IEEEauthorblockA{ \textit{George Mason University}\\
Fairfax, VA, USA \\
rpandey4@gmu.edu}
\and
\IEEEauthorblockN{Hemant Purohit}
\IEEEauthorblockA{ \textit{George Mason University}\\
Fairfax, VA, USA \\
hpurohit@gmu.edu}
\and
\IEEEauthorblockN{Bonnie Stabile}
\IEEEauthorblockA{ \textit{George Mason University}\\
Fairfax, VA, USA \\
bstabile@gmu.edu}
\and
\IEEEauthorblockN{Aubrey Grant}
\IEEEauthorblockA{ \textit{George Mason University}\\
Fairfax, VA, USA \\
agrant12@gmu.edu}
}
\maketitle

\begin{abstract}
 The recent surge in women reporting sexual assault and harassment (e.g., \#metoo campaign) has highlighted a long-standing societal crisis. This injustice is partly due to a culture of discrediting women who report such crimes and also, rape myths (e.g., `women lie about rape'). 
 Social web can facilitate the further proliferation of deceptive beliefs and culture of rape myths through intentional messaging by malicious actors.   
 
 This multidisciplinary study investigates Twitter posts related to sexual assaults and rape myths 
 for characterizing the types of malicious intent, which leads to the beliefs on discrediting women and rape myths. Specifically, we first propose a novel malicious intent typology for social media using the guidance of social construction theory from policy literature that includes \textit{Accusational}, \textit{Validational}, or \textit{Sensational} intent categories. We then present and evaluate a malicious intent classification model for a Twitter post using semantic features of the intent senses learned with the help of convolutional neural networks. Lastly, we analyze a Twitter dataset of four months using the intent classification model to study narrative contexts in which malicious intents are expressed and discuss their implications for gender violence policy design. 
\end{abstract} 

\begin{IEEEkeywords}
Intent Mining, Convolutional Neural Networks, Policy-affecting intent, Public Health Analytics, Rape Myths  
\end{IEEEkeywords} 


\section{Introduction}
\label{sec:intro}

Rape and sexual assault are pervasive, long-standing, societal problems.  One out of every six American women, and one in every 33 men - or about 17\% and 3\% of the population, respectively - have been the victim of an attempted or completed rape in their lifetime~\cite{rainn2017b}. Younger people are more likely to be victims of these crimes. In institutions of higher education (IHEs), 23.1\% of female and 5.4\% of male undergraduate students experience rape or sexual assault by use of physical force, violence, or incapacitation~\cite{cantor2015report}, yet an estimated 80\% of incidents are not reported~\cite{NAP18605}.  Rape or sexual assault are about half as likely to be reported to police as robbery (54\%) and aggravated assault (58\%), with the former being reported in only about a quarter of all cases (23\%)~\cite{bjs:2017}.  Social stigma surrounding sexual crimes likely contributes to this low level of reporting. This in turn likely emboldens perpetrators, who act with the confidence of relative impunity given the low level of reporting, and even more remote likelihood of prosecution. Rape and sexual assault constitute injustices that impose both human and financial costs on individuals, and society as a whole, while also furthering gender inequality. 

\begin{table*}[]
\centering
\caption{Anonymized example of messages with varied intent in the conversation on Twitter regarding sexual assault.}
\label{tab:intmsg}
\begin{tabular}{p{12cm}|c} 
\hline
\textbf{Twitter Message}   & \textbf{Policy-affecting Intent} \\
\hline
\textbf{\textit{M1}}.  white women have lied about rape against black men for generations                                                           & \textit{Accusational}            \\
\textbf{\textit{M2}}. Listening to \#Dutton say women on \#Nauru who have been raped are often lying makes me sick. Showing us once again his misogyny \&amp; sexism & \textit{Validational}            \\
\textbf{\textit{M3}}. There is no New Clinton, never has been. Shes the same rape defending, racist, homophobic liar shes been for 70 yrs URL                        & \textit{Sensational}    \\
\hline
\end{tabular}
\vskip -0.1in
\end{table*}

It is the role of law and policy to address public problems such as the mitigation of these sexual assault crimes out of an obligation to lessen harms, protect the autonomy of citizens, and secure justice. Public attitudes such as those reflected in social media allow the better understanding of the nature of public beliefs and the associated intentions at large scale~\cite{purohit2016gender}. Thus, in principle, social media mining can inform policymakers to ultimately help in the policy formulation and revision of laws as well as improve the policy outcomes. 
This research investigates prominent intentional message themes on Twitter regarding sexual assault in order to understand the extent and help mitigate the effect of rape myths - especially the myth that `women lie about rape', which is one of the most frequently endorsed rape myths~\cite{franiuk2008prevalence}. Using the guidance of \textit{social construction theory}, which explains how policy is influenced by perceptions of target populations~\cite{schneider2014democratic}, we propose a novel malicious intent typology and an automated classifier for categorizing Twitter messages into the relevant classes of \textit{Accusational} (blaming someone or a group), \textit{Validational} (endorsing a belief), and \textit{Sensational} (creating uncertainty and fear). 
Table~\ref{tab:intmsg} shows examples of such intentional messages. Addressing the problem of identifying such intentional messages related to sexual assault and rape on social media provides the intelligence to better understand and assess the context in which the public expresses deceptive beliefs. 
The specific contributions of this study are the following: 
\begin{enumerate}
\item We propose a novel malicious intent typology and an intent classification method using distributional semantics for social media messages. Our evaluation against several baselines shows the effectiveness of our feature representation of intent senses learned from convolutional neural networks. 
\item To our knowledge, this is the first large-scale multidisciplinary study to explore policy-affecting malicious intent and their contexts in social media, using a novel application of social construction theory. 
We present a scalable alternative for collecting information to help policy analysts for gender inequality and complement the costly survey-driven methods. 
\item We present novel insights on the context of malicious intents regarding sexual assault and rape myths in Twitter dataset collected over 4 months. We found that \textit{accusational} intent messages are the most prevalent in social media. Such messages reflect public beliefs that undermine the credibility of women who report rape and express more concern for accusers than the accused, with clear implications for policy debate, design{\color{blue},} and outcomes.     
\end{enumerate}

The rest of the paper is organized as follows. 
Section II provides a background on social construction theory and its novel application to understanding policy-relevant beliefs about sexual assaults including rape myths. 
Section III presents our approach for acquiring meaningful intent categories and classifying intentional messages. 
Section IV then describes the experimental setup with several baselines to classify intentional messages on Twitter and compare against the proposed model. Section V presents result analysis for categorization by intent typology as well as the topic modeling and psycholinguistics analysis of the context of intentional messages before the discussion in Section VI and conclusion. 

\section{Background and Related Work}
\label{sec:related}
This section presents related work on social construction theory, its use for identifying policy-affecting intent categories, and then finally, user intent modeling on social media.  

\subsection{Social Construction, Rape Myths, and Policy}

The social construction theory of target populations  explains ``who benefits and loses from policy change,'' depending on whether they are seen positively in the public sphere~\cite{schneider2014democratic}.  Those who are viewed in a negative light are less likely to find policies shaped in their favor, while those who are positively socially constructed and powerful (known as the ``advantaged'') are more likely to be benefited by policy.  Groups that are negatively socially constructed and weak (known as the ``deviants'') 
are more likely to be condemned in public discourse and disadvantaged by policy.  

\begin{table*}[]
\centering
\caption{Social construction framework and policy-relevant characterizations of actors in rape and sexual assault.}
\label{tab:framework}
\begin{tabular}{|C{1cm}|C{1cm}|C{7.5cm}|C{7.5cm}|}
\hline
\multicolumn{2}{l}{\multirow{2}{*}{}}                                                & \multicolumn{2}{c}{\textit{Social Construction}}                                                                                                                                                                                                                                       \\ 
\multicolumn{2}{l}{}      & \multicolumn{1}{c}{{\ul Positive}}      & \multicolumn{1}{c}{{\ul Negative}}           
 \\ \cmidrule(l){3-4} 
 
 \multicolumn{1}{l}{}     &  \multicolumn{1}{c|}{{\multirow{3}{*}{{\ul Strong}}}} & \multicolumn{1}{c|}{\textbf{Advantaged}}    & \multicolumn{1}{c|}{\textbf{Contender}}                                                                                                              \\ 
                       
    \multicolumn{1}{l}{\multirow{5}{*}{\textit{Power}}}    & \multicolumn{1}{l|}{}        & Accused cast as athletes, breadwinners, men with potential                                                             & Accusers have political power to influence policy, but may be cast negatively as promiscuous, feminist or abrasive                                   \\
   \multicolumn{1}{l}{}                      & \multicolumn{1}{l|}{}                                       & \textit{[Policy benefits the Accused, burdens Accuser]}                 & \textit{[Policy may move toward accountability for Accused]}                                                                                              \\ \cmidrule(l){3-4} 
                       
  \multicolumn{1}{l}{}                       & \multicolumn{1}{c|}{{\multirow{3}{*}{{\ul Weak}}}}   & \multicolumn{1}{c|}{\textbf{Dependents}}       & \multicolumn{1}{c|}{\textbf{Deviants}}                                                                                                               \\  
   \multicolumn{1}{l}{}          & \multicolumn{1}{l|}{}                                       & Accusers seen as innocents, victims, not blameworthy, but lack political power; Accused seen as premeditating criminals & Accusers cast as liars, ``sluts'' or vengeful women                                                                                                    \\
   \multicolumn{1}{l}{}         & \multicolumn{1}{l|}{}                                       & \textit{[Policy may benefit Accusers or move towards holding Accused more accountable]}                                     & \textit{[Policy benefits the Accused, burdens Accusers with vocal support of politicians and public]} \\
          \hline
\end{tabular}
\vskip -0.1in
\end{table*}

	Sexual assault policies primarily affect two populations: accusers (victims) and the accused (perpetrators).  Policies that facilitate the reporting and punishment of sexual assault benefit accusers, but are seen by some as infringing on the rights of the accused.  Some fear that promulgating such policies will result in an unreasonably high number of false reports of rape or sexual assault, while, quite to the contrary, evidence demonstrates that this has been a historically underreported crime. Criminal justice data indicate that the rate of false rape accusations is no more than false allegations of other criminal offenses~\cite{rumney2006false,edwards2011rape,gunby2013regretting}, and place false allegations of rape at around just 5\%~\cite{ferguson2016assessing}. Despite this low figure, public dialogue and policy discussions suggest that there is a belief that the incidence of false reports of rape is much higher. This false belief leads to policy outcomes that favor the rights and wellbeing of the accused over those of the accuser, and, paradoxically, may even contribute to the continued suppression of rape reporting. 

	Negative characterizations of accusers are evident in social constructions of women perpetuated through social media, among other means.  Feminist scholars have long argued that rape myths contribute to such characterizations by casting doubt on the very existence of rape, and that the widespread acceptance of rape myths has practical implications~\cite{burt1981rape}.  The Illinois Rape Myth Acceptance Scale, a tool used to measure rape myth acceptance~\cite{mcmahon2011updated} 
	is divided into four categories or subscales: 1) She asked for it; 2) He didn't mean to; 3) It wasn't really rape; and, 4) She lied. Taken together, these myths characterize women negatively, as lacking credibility, at a minimum, or even as routinely and willfully practicing deceit.  Such ideas are rooted in  long-standing cultural norms; the origins of the myth that women lie about rape as vengeful retaliation towards men who reject their advances can be traced back to Greek and Judeo-Christian theology~\cite{edwards2011rape}.  
	
Table~\ref{tab:framework} summarizes the policy-relevant characterization of the key actors in rape and sexual assault context for our analysis. In short, accusers (mostly young women) have historically been widely characterized negatively as lying or promiscuous - ``deviants'' in the social construction framework. They have also traditionally been cast as ``dependents,'' lacking political or economic power.  The accused (mostly young men) are often seen in terms of their prowess as athletes or students, or their promise as breadwinners, and so are positively constructed as ``advantaged''.  
	Advantaged groups are least likely to experience burdens and Deviant groups are at the highest risk, since ``punishment'' of those in this latter group ``yields substantial political payoffs'' for policymakers and political actors, as does rewarding those in the former group~\cite{schneider1997policy,bonnie2016reproductive}. Therefore, groups characterized as Advantaged and Deviant are the most discussed in political dialogue, a pattern that we expect to see mirrored in social media, where the payoffs might be counted in spreading of a deceptive idea or belief.  
	
	  Social media can be said to both reflect and perpetuate prevailing social constructions through both informal online dialogue and intentional messaging by various stakeholders. It has been demonstrated that rape myth acceptance is associated with negative attitudes about women~\cite{baugher2010rape}, stronger anti-victim, pro-defendant judgments~\cite{sussenbach2017looking}, and influencing ``what is considered a `legitimate rape' and who is considered a `credible victim'~\cite{brownmiller2013against}.'' Therefore, advancing understanding of social media's role in propagating (or combating) rape myths is expected to assist policymakers. Analyzing the context of malicious intent in propagating the rape myths, or otherwise deceptive beliefs to discredit women is the first essential step.    
	  
	    	We discuss some specific examples of policy that can precipitate and be influenced by social media dialogue on rape and sexual assault as follows. The federal guidance on Title IX issued by the Obama Administration and state level affirmative consent laws, are two classes of policy interventions designed to address the injustice of widespread campus sexual assault, its underreporting, and inadequate institutional response. The former, the U.S. Department of Education's Office for Civil Rights' (OCR) Dear Colleague Letter of 2011, articulated to all school districts, colleges, and universities that Title IX of the Education Amendments of 1972 would now consider ``sexual harassment of students, which includes acts of sexual violence…[as] a form of sex discrimination prohibited by Title IX'' of the Civil Rights Act of 1964. Likewise, California's ``Yes Means Yes'' law, passed in 2014, New York's ``Enough is Enough'' law and Illinois' ``Preventing Sexual Violence in Higher Education Act,'' passed in 2015, as well as Connecticut's ``Yes Means Yes'' law, passed in 2016, are another set of examples to include parallel provisions. Both the Obama Administration's Title IX guidance and state-level Affirmative Consent laws can be construed as benefiting victims and burdening the accused relative to their former positions. 
	    	
	The social construction framework shown in Table~\ref{tab:framework}, which explains how powerful stereotypes influence policy outcomes, can aid in understanding the contexts of the forces at play. For instance, in the opposition to existing policies and showing beliefs for rape myths by sharing intentional messages. 

\subsection{Modeling User Intent on Social Media}

User intent mining has been traditionally investigated in the domain of Information Retrieval and Web Search for better understanding of user query intent. The approach was to exploit historical user data from  search logs and click sequence graphs for broad categories of navigational, informational, and transactional intent~\cite{jansen2008determining}. It has been a relatively new area of investigation in social media research~\cite{purohit2018intent}, where recent works have investigated the intent classification problem for social media text in specific domains such as buying-selling intent for commercial products~\cite{hollerit2013towards} and help seeking-offering intent during disasters~\cite{purohit2015intent} as well as in general, across different topics for recommendations such as travel and food~\cite{wang2015mining}.          

	Intent classification is a special type of text classification focused on action-oriented cues in the text, which differs from topic classification, which is focused on the subject matter and sentiment or emotion classification, which is focused on the current state of affairs~\cite{kroll2009analyzing,chen2013identifying}. Therefore, the focus of feature representation and algorithms for learning have variations across these different types of problems.  
	
	Intent expressed in social media messages can be observed in both implicit and explicit forms. Implicit intent refers to latent or hidden aim for the action behind the expressed cues, such as message M2 in Table~\ref{tab:intmsg}, where the author is trying to validate the presented fact for the purpose of convincing others. In contrast, explicit intent refers to a specific aim for the action expressed in the text, for instance, message M1 in Table~\ref{tab:intmsg} clearly expresses the belief to accuse a particular gender group. Thus, our study requires the modeling of both explicit and implicit intent forms from short-text messages of social media, in contrast to the prior work on explicit intent identification in general~\cite{wang2015mining,chen2013identifying}. Next, we describe 
    our method to acquire relevant intent typology and then categorize messages for the intent types. 

\vskip -0.3in
\section{Approach: Policy-affecting Intent Analytics}    
\label{sec:approach} 

This section first describes our dataset and then the solution of an intent typology, followed by our classification method. 

\subsection{Dataset}

We collected Twitter posts using the keyword-based (`filter/track') method of Twitter Streaming API for the period of four months - August 1 to December 1 2016 using CitizenHelper system~\cite{karuna2017citizenhelper}. and the seed keywords of `rape' and `sexual assault'. Our dataset contained a total of 5,434,784 tweets. For our study, we created a subset of data containing deception or myth related terms using the following lexicon, chosen based on the prior literature about rape myths and reporting on sexual assault~\cite{franiuk2008prevalence,aiken1993false}: {\textit{lie, lying, lied, liar, hoax, fake, false, fabricated, made up}}. The filtered subset contained 112,369 tweets (referred as `myth dataset' in the paper for clarity), which we investigate for  policy-affecting intent. 

\subsection{Policy-affecting Intent Typology}

Using social construction theory and the constructed analytical framework as shown in Table~\ref{tab:framework}, we inferred the commonalities for potential malicious intent types across the four quadrants of policy-affecting characterization of actors. We list the prevalent intent types in the following. We also consulted two sexual assault policy subject-matter experts, who further independently reviewed and validated the policy-affecting intent types and the associated themes in the top 100 `retweet' messages (forwarded tweets) extracted from our myth dataset. The resulting specific intent categories are as follows: \textit{Accusational}, \textit{Validational}, \textit{Sensational}, or \textit{None}, where   
\begin{itemize}

\item	``\textit{\textbf{Accusational}}'' messages express doubts about or undermine accusers; express more concern for the accused than the accuser; and/or perpetuate the idea that women lie about rape. 

\item	``\textit{\textbf{Validational}}'' messages express belief in the accuser; and /or point out the injustice of the crime for an accuser or accused, and/or the inadequacy of the punishment. 

\item ``\textit{\textbf{Sensational}}'' messages focus more on politics or provocation than on the issue of rape or sexual assault; intent may be primarily to frighten, politicize or sensationalize with these terms, but not to affirm, accuse or meaningfully inform regarding rape or sexual assault. 

\end{itemize}

Messages categorized as \textit{Accusational} are expected to  both reflect and perpetuate the Advantaged status of the accused and the Deviant status of accusers. While \textit{Validational} messages might contribute to changing the quadrant category of the accused or accusers. For instance, accusers from the position of Deviants to a quadrant where they might be more positively socially constructed (as Dependents) or seen as having stronger political power (as Contenders).

\subsection{Intent Classification}
	We used the three key intent types to define our task of automated intent categorization in the following. 

Intent classification is a complex problem due to the likelihood of various intentional senses in a text that complicate natural language interpretation.  
	There are two key challenges for classifying intentional messages on the social media. First, the cues for indicating an intent category are sparsely present in a short-text message with the lack of sufficient contextual details. Thus, feature extraction to efficient capture intent representation becomes challenging due to the surrounding noise, resulting in poor learning of the useful regularities and patterns. Second, a specific intent can be expressed in a variety of textual forms (e.g., consider message M2 in Table~\ref{tab:intmsg} for the \textit{Validational} class illustration that could be written in different ways). Collecting such varied examples of intent expressions within each category to create a large training sample for efficient machine learning is a daunting challenge. Our proposed approach addresses these challenges of inferring intent from the short-text by observing an analogy in the domain of computer vision, where an image is constituted of multiple sparse cues that give the image an overall meaning and context. 
    Computer vision research has exploited a distributed semantic representation of the cues and applying deep learning methods for efficient performance in image processing tasks, such as Convolutional Neural Network (CNN). We, therefore, adapt the distributed semantic representation via word embedding for our feature representation of sparse intent cues in the short-text messages. However, for small dataset the CNN based learning approach is not efficient and generally, the deep learning approaches require large amount of training data for the requirement of optimizing a large number of parameters. Thus, we resolve to rather leverage the fully connected layer in the CNN architecture as an efficient feature representation in the traditional logistic regression classification model. We will discuss details and comparison with several baseline models in the following subsection after data annotation. \\ 

\subsubsection{Data Annotation}  
We randomly sampled 2500 unique messages from the myth dataset for annotation of intent categories. We asked for minimum three annotators to label each message with the four intent classes: \textit{Accusational}, \textit{Validational}, \textit{Sensational}, or \textit{None}. We provided instructions to label with five examples of each class.  For the resulting annotated messages, we used the confidence score greater than $67\%$ for finalizing the label of a message as per the annotation quality metric of Figure Eight crowdsourcing platform\footnote{https://success.figure-eight.com/hc/en-us/articles/201855939-How-to-Calculate-a-Confidence-Score}. 
We concluded with the final labeled dataset of 1163 messages. The resulting label distribution of the classes is: \textit{Accusational}:  530  (46\%), \textit{Validational}:  161  (14\%), \textit{Sensational}:  347  (30\%), and \textit{Other}: 125  (10\%).   

\begin{figure*}
\centering
\includegraphics[trim={0 3cm 0 5cm}, width=6in]{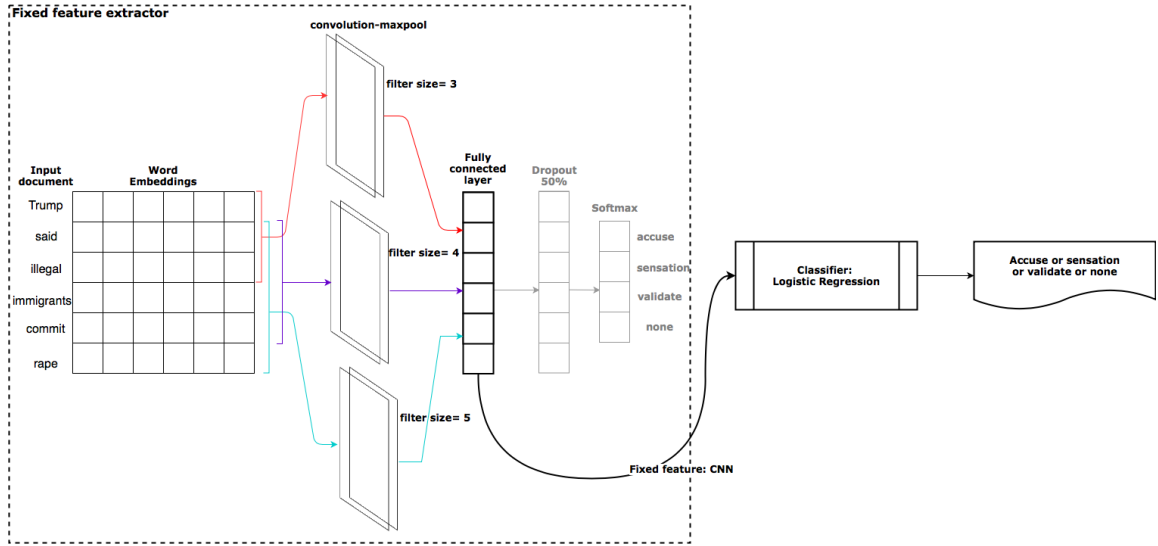}  
\caption{Summary of the proposed intent classification model that uses distributed semantics features as the last layer of CNN architecture.} 
\label{fig:model} 
\end{figure*}

The annotation results suggest an imbalance distribution of the policy-affecting intent messages, consistent with the real world machine learning tasks. 
 Next, we describe an automated multiclass intent classification approach to categorize the larger myth dataset for our analysis. \\   

\subsubsection{Feature Extraction and Learning Algorithm}

Prior research on intent classification on social media has used a variety of human-created rules as features~\cite{carlos2012intention,purohit2015intent} as well as automatically extracted features such as bag-of-words, n-grams, and Part-of-Speech tags~\cite{hollerit2013towards}. Instead of exploiting human-created rules as features due to the cost of creating an exhaustive list of rules to capture different intent expressions, we resolve to automatically generating the higher abstract-level features. We compare the classic Bag-of-Words model for features against the distributional semantics based model of word embedding for features, of the message text. In particular, we use the publicly available pretrained word2vec vectors for creating distributional semantic representation of each word in a message text. The pre-trained word vectors were generated by neural language models trained over 100 billion words of Google News data~\cite{mikolov2013distributed}. Prior to that, we used standard text cleaning for removing special characters, and only kept the words with minimum frequency of 3 (based on multiple trials.)  

We considered two different learning algorithms for our experiments. First, using a generalized linear model of logistic regression and second, using deep neural network model of CNN with softmax as classifier. The reason for exploring in-depth various feature-model combinations and the two algorithmic approaches 
is twofold. First, our dataset is small enough to train a model for efficient semantic representation by itself; hence, a pure deep learning based softmax classifier 
could not classify efficiently. Second, we have sparse cues for indicating intent that could limit the efficiency of feature representation by only the traditional Bag-of-Words or rule-based feature representation with linear models. 

Our resultant approach is to use the deep neural network model as the feature extractor with a traditional classifier. 
Specifically, the proposed model includes CNN codes (with adapted word2vec embedding) as features for a logistic regression classifier. We used a three step process as follows. First, we trained a CNN model, where we adapted the pretrained word2vec embedding to our event dataset for initialization, likewise~\cite{kim2014convolutional} (implementation details in baseline $B3$). Then, we used the fully connected layer output as the fixed feature set of CNN codes, 
which were used to train a 
logistic regression classifier. Figure \ref{fig:model} summarizes our proposed approach.  

\begin{figure}
\centering
\includegraphics[trim={0 6cm 0 5cm}, width=\linewidth]{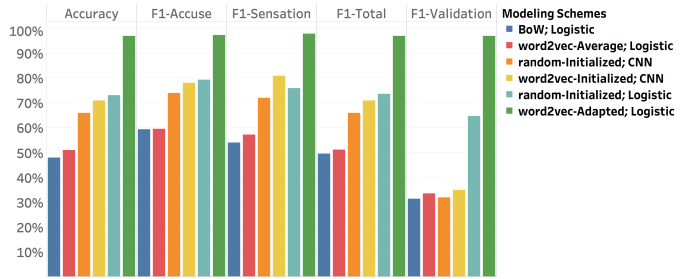} 
\caption{Comparison of stratified 10-fold CV results for different feature representation and machine learning models for policy-affecting intent classification. Our method `word2vec-adapted-Logistic' performs the best.} 
\label{fig:model-performance-graph} 
\vskip -0.2in
\end{figure}

\section{Experimental Setup}
\label{sec:experiments}

We evaluate the performance of our multiclass classification model using the standard metrics of accuracy and micro F-score with the stratified 10-fold cross validation (CV). Accuracy computes the percentage of the total number of correctly predicted messages, while micro F-score computes per class the value of weighted average of the precision (number of relevant messages among the predicted messages) and recall (sensitivity)~\cite{hall2011data}. 
We implement using python gensim library. We created the following baseline schemes: 
\begin{itemize}
\item \textbf{[$B1$] BoW Features + Linear Model}. 
We generated \textit{tf-idf} features for each message and trained with logistic regression model. 
We considered unigram words and then, applied Lovins stemming algorithm 
to get the stem words, and took maximum 1000 words as default parameters. 
\item \textbf{[$B2$] word2vec Average Features + Linear Model}. 
We computed the average word vectors of all the words in a message by using the pretrained Google's word2vec embeddings. Using the average vectors as features, we trained a logistic regression model.  
\item \textbf{[$B3$] Random Initialized Embedding + CNN (Softmax)}. 
We followed the approach of Kim~\cite{kim2014convolutional} for designing a CNN model for text classification. It consists of embedding of shape vocabulary size (735) x dimension (300) followed by 3 parallel convolutional-maxpool paired layer with convolution filter sizes as 3, 4 and 5 respectively. They are linked to a fully connected layer of size 384. Then, a dropout layer with 50\% dropout for regularization is connected and at last, softmax layer for classification probability. We added $l2$-loss for non-linearity and cross-entropy as loss function, which is reduced by Adam Optimizer~\cite{kim2014convolutional}. Training was done with higher learning rate (0.005) initially for faster reduction of loss, and then, we slowly decreased the learning rate up to 0.0001 to get the minimum loss. 
\item \textbf{[$B4$] word2vec Initialized Embedding + CNN (Softmax)}. 
We used the same architecture as above but initialized the embedding layers with Google's word2vec and then trained in the same way. 
\item \textbf{[$B5$] CNN (Randomly Initialized) Codes + Linear Model}. 
In this scenario, we used a two step process. First, we trained the CNN model same as the above experiments with our training data. Then, we used the fully connected layer output as the fixed feature vectors of CNN Codes. Finally, we used these CNN codes for training logistic regression classifier. 
\end{itemize}

\noindent Figure \ref{fig:model-performance-graph} shows the stratified 10-fold cross validation results. 

\section{Result Analysis}
\label{sec:results}

In this section, we first discuss the automated classification performance and intent prediction on the myth dataset using the proposed classifier. We then present an analysis and policy implication of the topical contexts of messages in the myth dataset categorized by the predicted intent classes.  
 
\subsection{Classifier Performance and Prediction Results}

Figure \ref{fig:model-performance-graph} shows the better performance of our proposed classifier in comparison to all the baselines, by leveraging the semantic representation based on CNN codes initialized with word2vec and trained on the generalized linear model of logistic regression. The external knowledge in word2vec of representing varied contexts for the words helps in addressing the challenge of sparsity in efficiently learning intent representations from a small training set. Modeling intent from social media text is a challenging task in general, such as 
maximum F-score of 58\% for classifying general intent types (e.g., travel, food, commercial goods, event) on Twitter~\cite{wang2015mining}. Albeit, our results show better performance of the proposed method for the domain-specific intent inference task. In particular, our approach achieved micro F-score for the policy-affecting intent classes up to 97.3\% for \textit{Accusational}, 96.9\% for \textit{Validational} and 97.8\% for \textit{Sensational} as well as both accuracy and macro F-score as 96.9\%. A potential reason for the best performance in learning \textit{Accusational} intent category is likely the explicit nature of intent expression when accusing a target. On the contrary, we note the inferior performance for the \textit{Validational} class in contrast to other classes potentially due to the highly implicit nature of \textit{Validational} intent expression as described earlier for the example M2 in Table~\ref{tab:intmsg}. We also intentionally did not balance the training dataset for learning to capture the real data distribution, as we apply the classifier for predicting categorical messages in the myth dataset next.

For our analysis, we consider only the unique messages for prediction task input (i.e. removing duplicates using Levenshtein string similarity $\geq$ 0.8, such as retweets or opinions on news headlines shared as tweets). The resulting non-duplicate message set comprised of 31,129 tweets and the intent category distribution of the predicted myth dataset is shown in Figure 3. 

\begin{figure}
\label{fig:model-performance}
\centering
\includegraphics[trim={0 8cm 0 7cm}, width=2.5in]{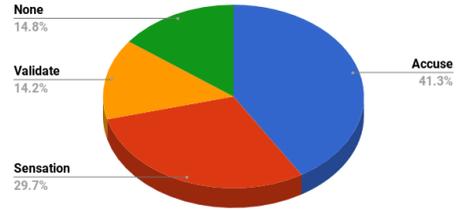} 
\caption{Distribution of predicted messages across the intent categories.}  
\vskip -0.2in
\end{figure}

	We observed the prevalence of \textit{Accusational} intent messages in the myth-related dataset. Examples of such predicted messages include the following, where feminists are being accused: 
	
\textit{``still a mra favourite tweet to feminists , just hoping that we will fire back w stats on false rape reporting url''}

It indicates the utility of social media as a source to collect information on public beliefs regarding sexual assault policy related actors - \textit{accusers} and \textit{accused} as well as stakeholders.  Furthermore, nearly one third of the messages are categorized as \textit{Sensational} indicating the role of social media as a channel to propagate agendas while mixing the context with any actor in 
the social construction framework, such as the following predicted message tries to create a narrative with political motives. 

\textit{``bill clinton who is been impeached, disbarred, accused of rape, other sexual misconduct, lying under oath is about tell"  }

	Next, we analyze the topical context of how the public expresses  intent in the categorized messages and the associated policy implications. 

\begin{table*}
\centering
\small
\caption{Topical cluster of words extracted from messages across different policy-relevant intent categories.}
\label{tab:cluster}
\begin{tabular}{c|C{5cm}|C{5cm}|C{5cm}}
                 & \multicolumn{1}{c}{\textbf{Acccusational}}   & \multicolumn{1}{c}{\textbf{Sensational}}   & \multicolumn{1}{c}{\textbf{Validational}}                                                                                                          \\
\textit{TOPIC 1} & rape url raped lie false women lying case men girls time white made black rapes saint accused money rapist shit 
                 & url lie fake sexual lied steal made charges accused criminal trial called court taxes raping muslim defended abuse white real 
                 & fake victims accused victim reported girl case fuck proven called white claim culture life making falsely understand reason stories rapists \\ 
\hline
\textit{TOPIC 2} & man asaram police hoax victims found year allegation stone forced fined number fact revenge females delhi free media lot rolling 
				& rapes year kill donald corrupt cheat vote hoax muslims pedophile proven laughed benghazi war hate calls ass world shit 
                & liar claims stop thing police report accusation innocent raping evidence low guilty actual charges feel talking woman hard lies assume \\ 
\hline
\textit{TOPIC 3} & lied bapu filed assault allegations real accuser stop guilty report lives hate charge feminists lies attention assaulted derrick trial support  
				& rape lying liar trump hillary raped women bill clinton assault victim false victims case murder racist media man support time 
                & lying women assault lied girls true ppl child bad problem year makes wrong good things call hate world calling assaults 
\end{tabular}
\end{table*}

\subsection{Topical Context Analysis} 

We conducted a topic modeling analysis of the predicted intent category sets. Topic modeling provides a mechanism to discover latent themes based on semantic associations between words by creating clusters that represent abstract topics. Although it requires parameter tuning to generate human-interpretable topics. A popular approach for topic modeling is to employ Latent Dirichlet Allocation (LDA) algorithm proposed by~\cite{blei2003latent}. After standard text preprocessing steps of stop word removal and lemmatization, we trained the LDA model with default parameters using the topic modeling implementation in MALLET toolkit~\cite{mccallum2002mallet}. 
We experimented for different number of topics and found 5 topics suitable for our analysis with distinct themes. Table~\ref{tab:cluster} shows the comparative analysis of the top three identified topical clusters (due to space limitation).   
	We observe that messages with \textit{Accusational} intent about rape and sexual assault myths have the context of intent about specific target groups across gender (`men' and `women'), race (`white'), religion (`asaram' is a religious leader), and occupations (`police') as evident. In contrast, messages with \textit{Sensational} intent focus on the trending news topics related to politics and current affairs, which is plausible with our theory that the prime purpose of such user intent is to create the alternative narrative for a target (e.g., a political actor `donald') in connection with the context of rape and sexual assault. We can observe a different theme for the \textit{Validational} intent messages, where the context is focused on verifying or validating the facts and stories on existing or alleged crimes against women (e.g.,  `reported', `claim', and `proven' in the Validational column).  

\subsection{Psycholinguistic Analysis}
We investigate the different psychological patterns in the context of expressing specific intent. For this purpose, we use the popular psychometric analysis software Linguistic Inquiry Word Count (LIWC)~\cite{pennebaker2015development}. LIWC counts words in psychologically meaningful categories for a given text. For this analysis, we randomly sampled 2000 messages from each of the predicted intent category set and ran through LIWC. 

	We found three key observations in the LIWC measures. First, \textit{Accusational} and \textit{Validational} intent messages use causal writing style, which can be associated with their objective of convincing others about different myths and claims of a rational approach. Second, \textit{Validational} intent messages express greater certainty in the context of communication, partly attributed to the observation in the topical context analysis where we found that the context of such messages in often about the facts or stories from the past. Third, \textit{Sensational} intent messages use greater expression of power and negative emotion than any of the other categories. It can be understood in the context of creating sensation by bullying and showing strong subjectivity towards a target or topic. These observations provide guidance towards the design of potential features to improve identification of policy-affecting intent messages as well as their topical context in future studies.

\section{Discussion}
\label{sec:discuss}

This study presented novel insights on the contexts of using social networks as a means to reflect or perpetuate rape and sexual assault related myths. We investigated a novel adaption of the framework of social construction theory that helped identify policy-relevant actors in the conversations about rape and sexual assault on social media. Using the guidance of the social construction framework, we discovered three types of policy-affecting intent categories and proposed a novel intent categorization scheme for social media. This approach provides a scalable alternative for collecting information that can assist the policy analysts on gender violence and can complement the existing costly, survey-driven methods. 
We demonstrated a novel design of intent classifier for short-text 
by using a rich feature representation based on the adaptation of pretrained word2vec embedding, which helped overcome the challenge of efficiently capturing the sparse cues of intent indicators in text messages with large labeled data. 
Finally, we analyzed a four month data set to identify the themes in which intent is  expressed regarding myths about rape and sexual assault. We observed that \textit{Accusational} intent messages are the most prevalent in social media, with a contextual focus on public beliefs. They target the credibility of women and highlight the `advantaged' status of male accusers that has a clear influence on social construction and policy development.  

  This research study has, however, some limitations that provide a direction for future work. 
  We have only used English language tweets for a fixed time duration and thus, did not account for the time-based comparison of different gender myths events across multiple languages. It is partly due to the challenge of modeling intent in another language, where the semantics of how intents are expressed could require a different approach to develop an efficient feature representation for learning. Also, since the data collection was done based on keyword-based approach, in future one can study the presence of malicious intent types in the dataset collected through random samples of tweet stream for a particular location. Given the complexity of modeling policy-affecting intent, we also plan to investigate novel neural language models 
  while incorporating perceptual and cognitive features associated with specific intent types, as identified in the LIWC analysis. Finally, the presented analysis approach using the social construction framework can be extended for studying rape and sexual assault related myths on another social networking platform. Also, a future study could compare the prevalence of specific policy-affecting intent categories and their context across the different social networking platforms.      
\section{Conclusions}
\label{sec:conclusion}

In this paper, we presented the first quantitative analysis of policy-affecting intent expressed on social media regarding rape and sexual assault by a novel application of social construction theory. We showed that by using a social construction framework, meaningful categories of 
malicious intent associated with public beliefs can be identified, with key policy design implications. 
We demonstrated a novel CNN-based policy-affecting malicious intent classifier with micro F-score up to 97\% for an intent class, by representing and learning the semantics of sparse intent cues in the short-text Twitter messages via external knowledge of word2vec embeddings. 
When compared with traditional methods based on bag-of-words representation, the deep learning based CNN approach with pre-trained word vector representations generated more optimal features for efficient learning. The identified intent-related messages were used to discover the contexts using topic modeling analysis, where we found that the public uses social media regarding rape and sexual assault extensively for \textit{Accusational} and \textit{Sensational} intent themes with a focus on targeted groups. Such targets include race and occupation that aim to undermine the credibility of women and highlight the Advantaged status of male accused. This analysis presents a direction for the use of social media analytics for assisting the information needs of policy analysts for the design, development, and analysis of rape and sexual assault related policies.   
\vskip 0.05in
\spara{Reproducibility.} \textit{Data is available upon request for research.} 
\section{Acknowledgement}
Authors thank WI'18 reviewers for valuable feedback and also, U.S. National Science Foundation for grant IIS-1657379. 

\bibliographystyle{IEEEtran}
\bibliography{paper-wi-gbv-intent.bib} 

\end{document}